\documentclass[preprint2,openany]{aastex7}

\shorttitle{Spectropolarimetry of GRB 181201A}
\shortauthors{Bal et al.}

\usepackage{graphicx}
\usepackage{float}
\usepackage{amsmath}
\usepackage{subfigure}
\usepackage{xcolor}
\usepackage{makecell}
\usepackage{multirow}
\usepackage{soul}
\usepackage{url}
\setstcolor{red}
\usepackage{gensymb}
\usepackage{hyperref}

\begin{document}

\title{GRB 181201A: Spectropolarimetric Constraints on Prompt Emission and Magnetic Field Structure}

\author[0009-0003-8022-8151]{Dhruv S. Bal}\email{dbal@clemson.edu}\thanks{Email: dbal@clemson.edu}
\affiliation{Department of Physics and Astronomy, Clemson University, Clemson, SC 29634, USA}

\author[0009-0002-2068-3411]{Nikita S. Khatiya}\email{nkhatiy@clemson.edu}
\affiliation{Department of Physics and Astronomy, Clemson University, Clemson, SC 29634, USA}

\author[0000-0001-9094-0335]{Philippe Laurent}\email{philippe.laurent@cea.fr}
\affiliation{Université Paris-Saclay, Université Paris Cité, CEA, CNRS, AIM, 91191, Gif-sur-Yvette,
France}

\author[0000-0001-9494-0981]{Diego G\"{o}tz}\email{diego.gotz@cea.fr}
\affiliation{Université Paris-Saclay, Université Paris Cité, CEA, CNRS, AIM, 91191, Gif-sur-Yvette,
France}

\author[0000-0002-8028-0991]{Dieter H. Hartmann}\email{hdieter@clemson.edu}
\affiliation{Department of Physics and Astronomy, Clemson University, Clemson, SC 29634, USA}

\author[0000-0003-0516-2968 ]{Ramandeep Gill}\email{rsgill.rg@gmail.com}
\affiliation{Astrophysics Research Center of the Open University (ARCO), The Open University of Israel, P.O Box 808, Ra’anana 43537, Israel}
\affiliation{Instituto de Radioastronomía y Astrofísica, Universidad Nacional Autónoma de México, Morelia, Michoacán, C.P. 58089, México}

\begin{abstract}
Gamma-ray bursts (GRBs) have served as testbeds for high-energy astrophysical phenomena since their discovery over 50 years ago. GRB detections reveal diverse properties across events, complicating efforts to constrain their emission mechanisms, magnetic fields, jet structures, and compositions. Spectropolarimetry is a powerful tool utilized to disentangle these properties. INTEGRAL/IBIS has previously detected three polarized GRBs 041219A ($65\pm 26\%$), 061122 ($>$33\%) and 140206A ($>$28\%). We present the time-integrated and time-resolved analyses of the fourth-polarized GRB 181201A. This GRB exhibits high polarization ($81.5^{+17.2}_{-15.7}\%$) over three time bins with the highest signal-to-noise ratio, and the polarization angle (PA) shows a modest temporal evolution. The photon index does not cross the synchrotron line-of-death and shows a hard-to-soft evolution. A high degree of polarization (PD) tends to disfavor a photospheric emission model, as previous studies indicate that this model yields PD $<$ 20\%. A modest evolution in the PA points towards a non-axisymmetric jet with a multi-patch emission. Our results indicate that under standard synchrotron conditions, the magnetic field is coherent but lacks global structure. Moreover, the field has multiple patches, but only one or a few dominate due to high degrees of polarization, which overlap with the intrinsic maximum of synchrotron emission within 3$\sigma$ errors. The high PD is also indicative of the Compton Drag scenario, but only under certain conditions, such as viewing the relativistic jet near its edge. However, given our error bars and the scope for improvement in our pipeline, we cannot strongly favor either scenario; both may remain viable.
\end{abstract}

\section{Introduction}
The field of GRB physics has advanced dramatically since its discovery by the Vela satellites \citep{1973ApJ...182L..85K}. While both theory and observations have revealed new insights into the emission mechanisms of GRBs, their magnetic field structure, and their jet structure, many aspects about the prompt emission remain unknown \citep{2004RvMP...76.1143P,2015SSRv..191..471G,2015PhR...561....1K}. 

Utilizing polarization data is essential for overcoming degeneracies in temporal and spectral analyses, as different emission mechanisms predict distinct polarization signatures \citep{2009ApJ...698.1042T,2020A&A...644A.124K,2021Galax...9...82G}. Combined with the spectral information, 
spectropolarimetry is a powerful tool for probing properties such as the dominant emission mechanism, the magnetic field structure, and the jet structure. Several studies in the literature have demonstrated that distinct polarization signatures may show up by changing field configurations and dominant emission mechanisms, using time-resolved and time-integrated analysis for axisymmetric jets \citep{2009ApJ...698.1042T,2014MNRAS.440.3292L,2014ApJ...789..159I,2020MNRAS.491.3343G,2021Galax...9...82G,2021MNRAS.504.1939G} and non-axisymmetric jets \citep{2024MNRAS.52712178G}. 

However, only a few instruments can detect the polarization of the GRB prompt emission. Since prompt emission is dominated by hard X-rays and soft gamma-rays, the primary mechanism for their detection is Compton scattering in the detector. Photons preferentially scatter perpendicular to their electric field vector, and robust measurements require sufficient photon statistics, which in turn requires a large effective area, event reconstruction capabilities, and well-calibrated instruments \citep{2016A&AT...29..205C,2017NewAR..76....1M,2021Galax...9...82G}. After the launch of INTEGRAL, the two-layer construction of IBIS and its event-reconstruction capabilities made it possible to use it as a prompt polarimeter. However, it was recently decommissioned in 2025 after spending more than two decades in space, providing vital science to the astrophysics community \citep{2007A&A...466..895M,2008ApJ...688L..29F,2025JHEAp..45...96K,2026A&A...705A...7S}. Furthermore, it is worth noting that the instrument was not calibrated on the ground for polarization, but a recent study by \citet{2026A&A...707A.340B} has presented an in-orbit calibration using 20 years of data for Crab. To date, INTEGRAL has reported three GRBs with polarization: GRBs 041219A, 061122, and 140206A \citep{2009ApJ...695L.208G,2013MNRAS.431.3550G,2014MNRAS.444.2776G}. They report a high degree of polarization for all three GRBs. GRBs 061122 ($>$33\%) and 140206A ($>$28\%) are reported as lower limits at a 90\% confidence interval while GRB 041219A had a polarization degree of $65\pm 26\%$. The authors conclude that high polarization is indicative of a synchrotron origin \citep{2019NewAR..8701537G}.  

\citet{2020A&A...644A.124K} indicate that the central values of 11 out of the 14 time-integrated GRBs in the POLAR catalog have low polarization degrees ($<20\%$) and the time-integrated results thus disfavor a synchrotron model with a globally ordered magnetic field configuration. However, they do not entirely disfavor synchrotron, as the low polarization may arise from a time-varying polarization angle.\citep{2020A&A...644A.124K}. While the discrepancy between POLAR and INTEGRAL GRBs is interesting, it will be difficult to resolve with current data, as there are no common GRBs, and neither instrument is currently operational. Similarly, using data from IKAROS aboard the Gamma-Ray Burst Polarimeter (GAP) \citep{2011PASJ...63..625Y}, their team has reported three polarized GRBs: 100826A, 110301A, and 110721A. They find that the polarization degree is high for GRBs 110301A ($70\pm22\%$) and 110721A ($84^{+16}_{-28}\%$) \citep{2012ApJ...758L...1Y} while low for GRB 100826A (($27\pm11\%$)), as the time-resolved analysis of the third revealed a 90$^\circ$ flip of the polarization angle resulting in a low polarization degree \citep{2011ApJ...743L..30Y}. Joint observations from future space observatories, such as POLAR-2 \citep{2021SPIE11444E..2VH} and the Compton Spectrometer and Imager (COSI) \citep{2019BAAS...51g..98T}, and Daksha \citep{2024ExA....57...24B}, as well as currently operational instruments like AstroSat-CZTI \citep{2017JApA...38...31B}, will be vital for resolving these discrepancies. These issues have prevented the GRB community from reaching consensus on the mechanism of GRB prompt emission.

While most GRBs are expected to produce a Band function \citep{1993ApJ...413..281B} spectra, the underlying physical mechanism that produces it remains to be clearly understood. Synchrotron-dominated and dissipation- or Comptonization-dominated models can give rise to the Band spectrum \citep{2015PhR...561....1K}. However, when these mechanisms are combined with magnetic field configurations and jet structure, they   
lead to unique polarization signatures \citep{2009ApJ...698.1042T,2021Galax...9...82G}. However, as we record more polarization data, we can populate the Toma plot \citep{2009ApJ...698.1042T} to discern the emission mechanism of various GRBs. Thus, utilizing the polarization capabilities of any available instrument remains vital. 

In this study, we present the analysis of the full INTEGRAL-IBIS GRB archive spanning from 2002 to 2024. After data cuts, we are left with one GRB, GRB 181201A, in addition to the three previously detected by INTEGRAL/IBIS and INTEGRAL/SPI. The paper thus focuses on the analysis and the physical interpretation of the GRB jet properties derived from our results for this GRB. The paper is organized as follows: Section \ref{sec:Methodology} describes data selection and analysis. Section \ref{sec:Integrated_analysis} contains the results of time-integrated polarization and spectral analysis. Section \ref{sec:Resolved_analysis}  contains the results of the time-resolved polarization.
Section \ref{sec:Discussion} discusses the results and their physical interpretation. We conclude the paper with Section \ref{sec:Conclusion}.

\section{Methodology}
\label{sec:Methodology}
\subsection{Data selection}
IBIS can be used to detect polarization via Compton scattering because it has two layers, ISGRI and PiCSiT \citep{2003A&A...411L.131U}, and it primarily operates in the 15 keV to 10 MeV range. Photons that hit both layers within a nearly coincident timestamp are labeled as ``Compton Events" and are the primary source for polarization studies, while those that are not coincident are ``Single Events" and are extensively utilized for temporal and spectral analysis.
We start by downloading the table provided on the ISDC website \footnote{\href{https://www.astro.unige.ch/integral/science/grb\#ISGRI}{ISGRI burst catalog}} for all INTEGRAL-IBIS bursts. We select GRBs within 7$\degree$ of the instrument boresight because they are expected to yield more counts. Furthermore, we also include any GRBs beyond the 7$\degree$ boresight with high Compton counts that are clearly distinguishable from the background. After shortlisting the GRBs that satisfy this cut, we use the archival catalog hosted by ISDC \footnote{\href{https://www.astro.unige.ch/integral/archive\#Browse}{INTEGRAL archival data}} to retrieve the data. 

Only GRBs with a high signal-to-noise (S/N) ratio in the Compton counts of the lightcurve are used for further analysis, as they provide reliable measurements. Apart from the three previously reported GRBs, 041219A, 061122 and 140206A, we find no other GRBs that satisfy both data cuts. GRB 181201A was included in the analysis, even if it was $\sim11^\circ$ off-axis, as it had a high S/N in the Compton count space. Since the analysis of the first three was completed in previous studies \citep {2009ApJ...695L.208G,2013MNRAS.431.3550G,2014MNRAS.444.2776G}, this paper focuses on GRB 181201A.




\subsection{Data analysis}

For the polarization analysis, using the same pipeline for previously reported GRBs 041219A, 061122, and 140206A, we analyze the data for GRB 182101A. We use the same method as \citet{2009ApJ...695L.208G}, originally developed to detect polarization in the Crab Nebula using INTEGRAL data \citep{2008ApJ...688L..29F}. 

The GRB pipeline works similarly: it is designed to reconstruct the event using energetics and to calculate the distribution of counts over azimuthal scattering angles. The azimuthal scattering angle distribution (ASAD) for the GRB is then fit with Equation \ref{eq:ASAD} and the best-fit is determined using $\chi^{2}$. Furthermore, the pipeline also checks the image signal-to-noise ratio (S/N) using Compton counts from IBIS. When the image's S/N is below 9, the polarization results are not significant. This is in line with the recent analysis by \citep{2026A&A...707A.340B} which used 6 as the limiting value, but we remain more conservative as an S/N of 9 is roughly 3$\sigma$. For time windows satisfying this condition, the polarigram is then fitted with a cosine wave defined by

\begin{equation}
\label{eq:ASAD}
    N(\phi)=S[1+a_0\ \cos\ 2(\phi-\phi_0)]
\end{equation}
Here, $N$ is the number of detected Compton events as a function of the azimuthal angle $\phi$, $S$ is the normalization, $a_0$ is the modulation amplitude, and $\phi_0$ is the direction of the maximum scattering. The polarization degree and angle can thus be derived by using $\rm{PD}=a_0/a_{100}$ and P.A$=\phi_0-\pi/2+n\pi$. All polarization information reported in this paper is between 200 keV and 800 keV.

The a$_{100}$ is the modulation factor when we assume all incoming photons have 100\% polarization degree. To calculate, we follow the same method as described by \citet{2008ApJ...688L..29F}. After performing Monte Carlo simulations coupled with GEANT4 simulations \citep{Agostinelli:2002hh}, we find that a$_{100}$ is 0.35 for a $\theta$ of 11$^\circ$. This value is fixed for all calculations of polarization degree for GRB 181201A, as it was also detected at an off-axis angle of $\sim11^\circ$. We further describe the pipeline in Appendix \ref{app:ibis_polarization_pipeline}.

For the spectral analysis, we utilize the Offline Science Analysis (OSA) v11.2 pipeline \citep{2003A&A...411L..53C}. This is a publicly available software pipeline on the INTEGRAL software page. The pipeline uses raw telemetry files to generate a corrected event list that accounts for reconstruction and count-correction steps. Since the pipeline accepts a ``Good Time Interval (GTI)" file, it enables us to perform a time-resolved analysis of the source. 

It is important to note that IBIS uses a coded-detector mask, so the background is modeled by fitting the contributions of all point sources in the field of view, including internal noise and diffuse sources. After background subtraction, the pipeline extracts temporal and spectral data based on the specified time and energy intervals. 

The IBIS/ISGRI spectral results nominally lie within the 20-500 keV range. However, we observe low count rates and higher errors beyond $\sim$200 keV; thus, our spectral analysis is restricted to 200 keV.  
Since the analysis range of spectral (20-200 keV) and polarization (200-800 keV) analysis differs, our spectral fit does not constrain the photon indices in the polarization band. Thus, we cannot directly test whether the maximum polarization degree predicted by synchrotron emission, which assumes a local index \citep{rybicki1985radiative}, is consistent with our results. While different emission components can contribute across different energy regimes, we cannot test this scenario because our energy ranges do not overlap.
\begin{figure}[ht]
    \centering
    \includegraphics[width=\columnwidth]{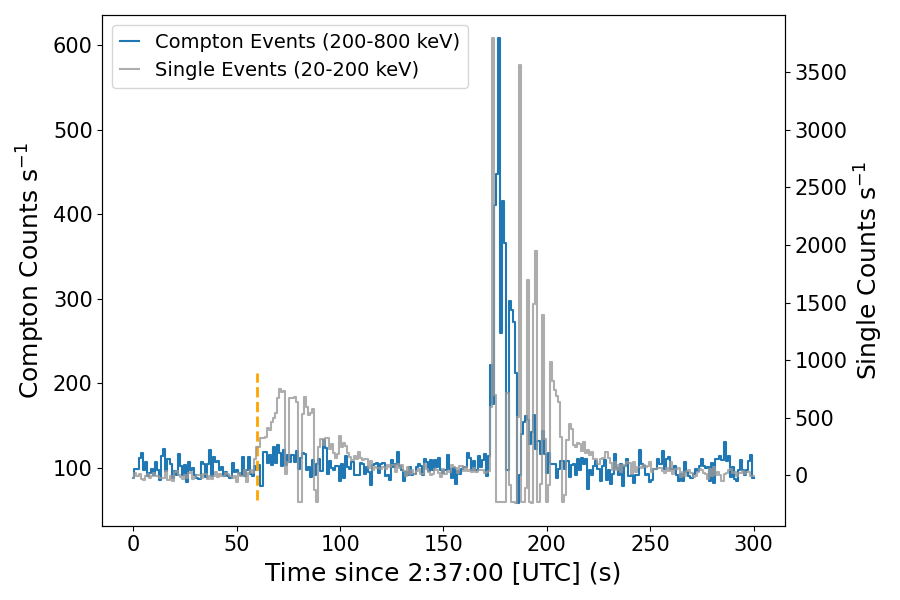}
    \caption{These are the lightcurves of single and Compton events detected in the science window. The gray line represents the former, while the blue line represents the latter. Similarly, the left y-axis is for Compton counts, while the right y-axis is for single counts. The orange line marks the GRB trigger in the IBIS catalog. The gaps in the lightcurves are due to telemetry limitations.}
    \label{fig:GRB_Comp_LC}
\end{figure}

\begin{figure*}[ht!]
    \centering
    \includegraphics[width=0.95\textwidth]{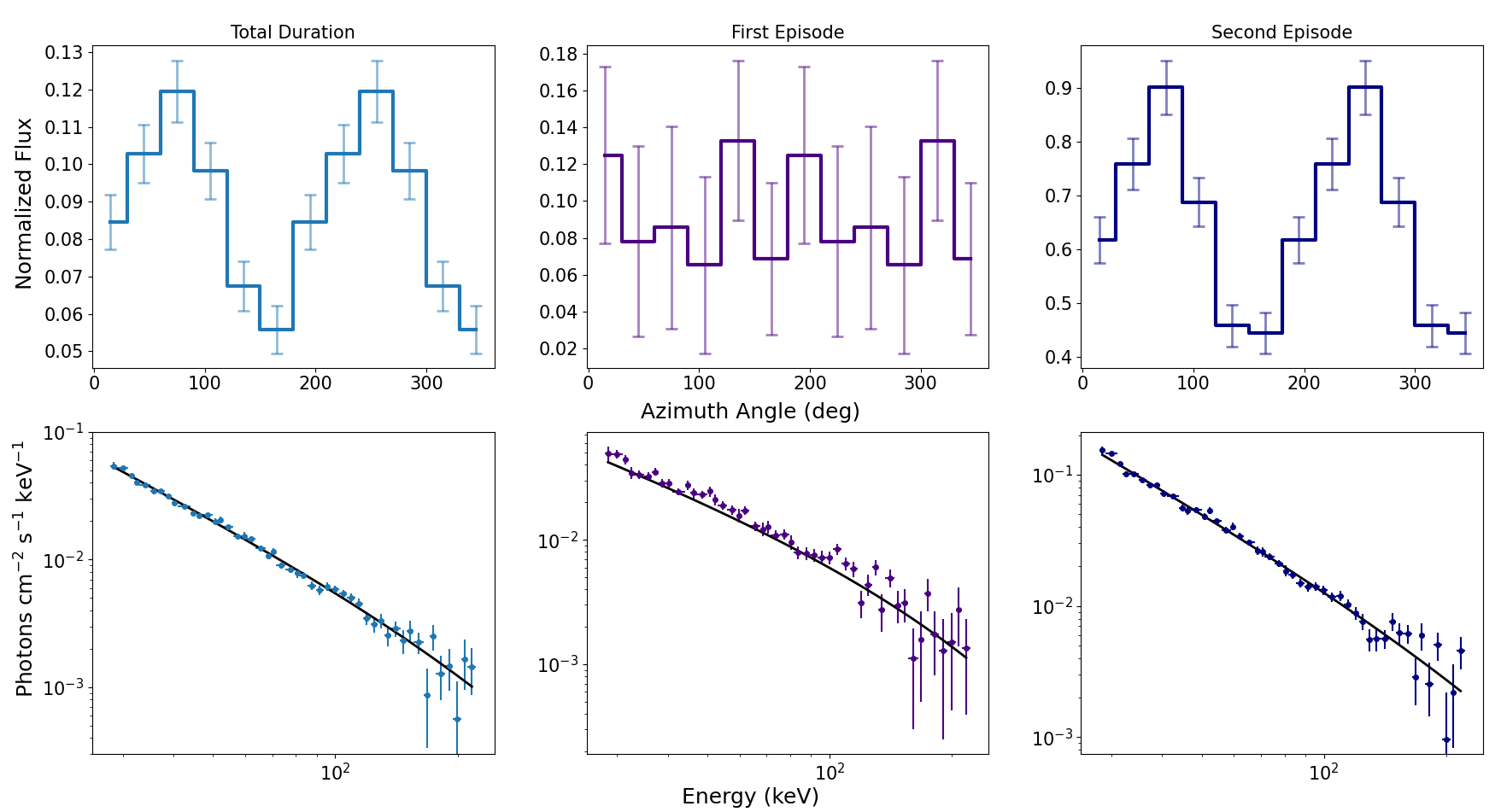}
    \caption{Top panels: The time-integrated azimuthal scattering angle distribution (ASAD) or polarigram of the GRB. Bottom panel: The time-integrated photon spectrum of the GRB. All spectra in this figure are best fit with a cutoff power law.}
    \label{fig:GRB_Pol_int}
\end{figure*}

\section{Time integrated and individual peak analysis}
\label{sec:Integrated_analysis}
\subsection{Polarization}
GRB 181201A was triggered at 2:38:00 UTC on December 1, 2018 \citep{2018GCN.23469....1M}. The trigger time is indicated by an orange vertical line in Figure \ref{fig:GRB_Comp_LC}. A slight rise and then a decrease are visible immediately after the trigger time and are coincident with the single-event light curve published on the ISGRI burst catalog page. Around 2:39:41 UTC, there is a stronger episode also coincident in both lightcurves. For the time-integrated analysis, we thus divide the burst into three time episodes. The full duration is 2:38:00 UTC - 2:40:36 UTC (full episode), the first emission episode is 2:38:00 UTC - 2:38:40 UTC, and the second emission episode is 2:39:41 UTC - 2:40:36 UTC. 

The polarization degree (PD) is 91$\pm$5 \% and 93$\pm$8\% for the full and second episode, respectively. Due to low S/N in the first episode, the degree and angle of polarization are not constrained. This indicates that the first emission episode does not contribute towards the polarization of the burst, as the error on the PD was very high, the signal-to-noise ratio was low, and thus PD remained unconstrained. The second emission episode is responsible for the burst's high PD. We thus focus the time-resolved analysis on episode 2, and any time-integrated values reported for this burst will also refer to episode 2 in this paper. We report the PD and PA for the three episodes in Table \ref{tab:time_int_pol} and display the results in the top panel of Figure \ref{fig:GRB_Pol_int}. 

\begin{figure*}
\begin{interactive}{js}{time_resolved_plotly_new.zip}
\includegraphics[width=0.9\textwidth]{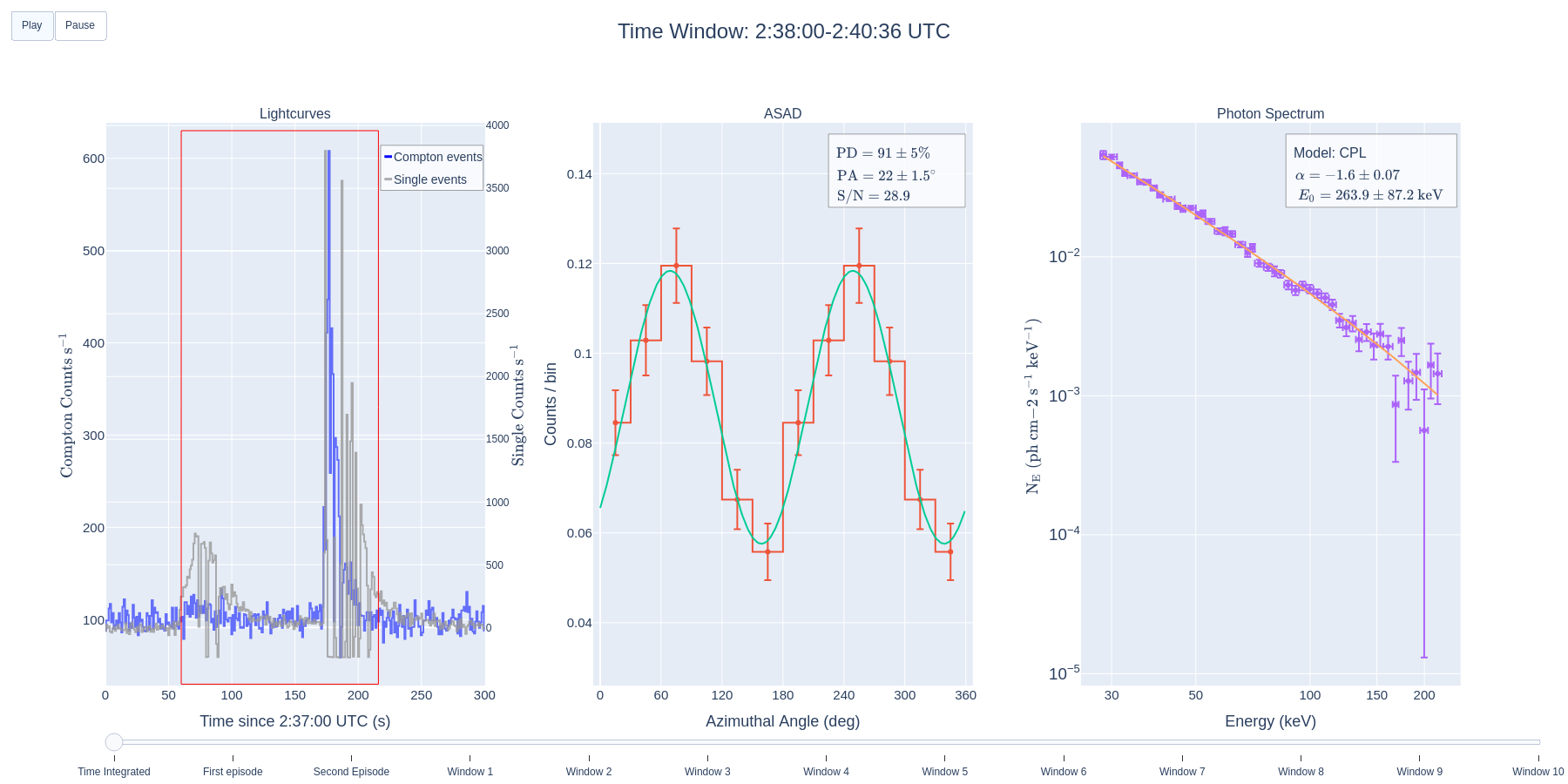}
\end{interactive}
\caption{Example figure of the spectropolarimetric dashboard. The title of the plot is the analysis for the time window of the three plots. The left panel indicates the Compton lightcurve, the single event lightcurve, and a red box highlighting the current analysis window. The middle panel shows the azimuthal scattering angle distributions for the analysis window, with the polarization degree, angle, and image Signal-to-Noise ratio displayed in the legend. The right panel shows the spectral analysis for the same time window, with the model parameters displayed in the legend in the top-right corner. The slider can be used to view different analysis windows, and the ``Play/Pause" button on the top left can be used to view the results as a movie. For best experience, please use the ``View in tab" option available in the Online article.} 
\label{fig:interactive}
\end{figure*}

\subsection{Spectral}
The spectral analysis is presented in the same time windows as the polarization analysis. We extract the data from the OSA pipeline and then pass it to XSPEC for further analysis \citep{1996ASPC..101...17A}. For all three windows, the best-fit spectral model was a Cutoff Powerlaw (CPL) model given by,

\begin{equation}
    N_E=A\left(\frac{E}{1\ {\rm keV}}\right)^{\alpha}e^{-\left(\frac{E}{E_0}\right)} \
\end{equation}

where $N_E$ is the photon flux in the units of $\mathrm{photons\ cm^{-2}\ s^{-1}\ keV^{-1}}$, $A$ is the normalization in the same units, $E$ is the energy in keV, $\alpha$ is the photon index, and $E_0$ is the characteristic energy also in keV. The peak energy is related to the characteristic energy by the equation $E_{\rm peak}=(2+\alpha)\times E_0$. 

The best fit model was selected based on a combination of the corrected Akaike Information Criterion ($\mathrm{AIC_c}$) and the $\chi^2$ value. If $\Delta\mathrm{AIC_c}$ between two models was greater than 5, then the model with the lower $\mathrm{AIC_c}$ was selected. However, if the difference was less than 5, the model with the lower reduced $\chi^2$ ($\chi^2$/degrees of freedom) was selected as the best fit. The reason for using $\mathrm{AIC_c}$ is that the ratio of the number of data points/ parameters is $<40$ \citep{burnham2002model}. The expression that we used is defined in \citep{Hurvich1989}

\begin{equation}
    AIC_c= \chi^2+2k+\frac{2k(k+1)}{N-k-1}
\end{equation}

where $k$ is the number of parameters in the model, and $N$ is the number of data points. The spectral characteristics are summarized in Table \ref{tab:time_int_spec} and displayed in the bottom panel of Figure \ref{fig:GRB_Pol_int}.

\section{Time resolved analysis}
\label{sec:Resolved_analysis}
\subsection{Polarization}
The second emission episode shows high Compton counts, enabling time-resolved polarization and spectral analyses. Consistent with the previous study by \citet{2009ApJ...695L.208G}, we select 10-second windows with 5-second overlap between 2:39:41 UTC and 2:40:36 UTC. The windows remain consistent for polarization and spectral analysis. We summarize the results in Table \ref{tab:time_res_pol} and in an interactive plot (Figure \ref{fig:interactive}). The evolution of the degree and angle of polarization is discussed in detail in the discussion section.

There are several time windows during which the error in the polarization degree was extremely high, and the polarization degree remained unconstrained. For such windows, we do not report the PD and PA Moreover, only values with an S/N ratio greater than 9 are reliable. Thus, any value below 9 is also reported as not defined, similar to the time-integrated analysis. Moreover, for time windows 6 and 7, we only report lower-limits. The procedure to calculate the same is outlined in \citet{2008ApJ...688L..29F,2013MNRAS.431.3550G,2014MNRAS.444.2776G}. The joint polarization degree and angle plots for these windows are included in the Appendix, and the reported values are within a 90\% confidence interval.

\subsection{Spectral}
OSA does not detect the GRB in the first window; therefore, we do not report any spectral information for it. In the other time windows, the spectral model evolves, and the best-fit model differs across them. For several time windows, the best-fit model is a power law described by

\begin{equation}
    N_E=A\left(\frac{E}{1\ {\rm keV}}\right)^{\alpha}
\end{equation}

where $N_E$ is the photon flux in the units of $\mathrm{photons\ cm^{-2}\ s^{-1}\ keV^{-1}}$, $A$ is the normalization in the same units, $E$ is the energy in keV, and $\alpha$ is the photon index. The results for spectral analysis are tabulated in Table \ref{tab:time_res_spec} and displayed in an interactive plot (Figure \ref{fig:interactive}). We note that \citet{2020AstL...46..783B} have used INTEGRAL/IBIS primarily for detection, timing, and localization constraints, but do not provide a spectral parameter fit using IBIS data. Furthermore, \citet{2019ApJ...884..121L} use Konus-Wind data \citep{2018GCN.23495....1S}, which have a wider energy range (20 keV-10 MeV); thus, their best-fit model is a Band spectrum, with best-fit values as $\alpha=-1.25\pm0.05$, $\beta=-2.73 ^{+0.11}_{-0.12}$, and E$_{\rm peak}=152\pm6$ keV, which differs from our models. 

Figure \ref{fig:interactive} is the ``Spectropolarimetric dashboard" containing information for the lightcurve in the left panel, the polarization information in the middle panel, and the spectral information in the right panel. For completeness, we also include the time-integrated analysis in this interactive plot, providing a single, comprehensive view across all time windows for GRB 181201A.

\begin{figure*}[ht!]
    \centering
    \includegraphics[width=0.45\textwidth]{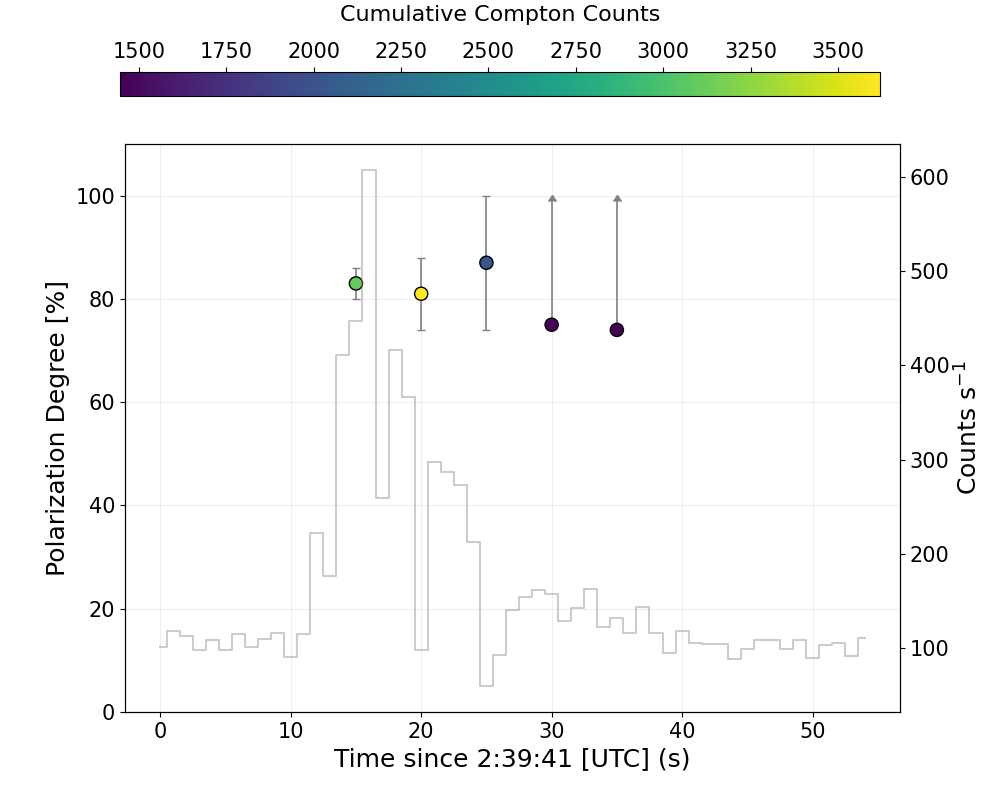}
    \includegraphics[width=0.45\textwidth]{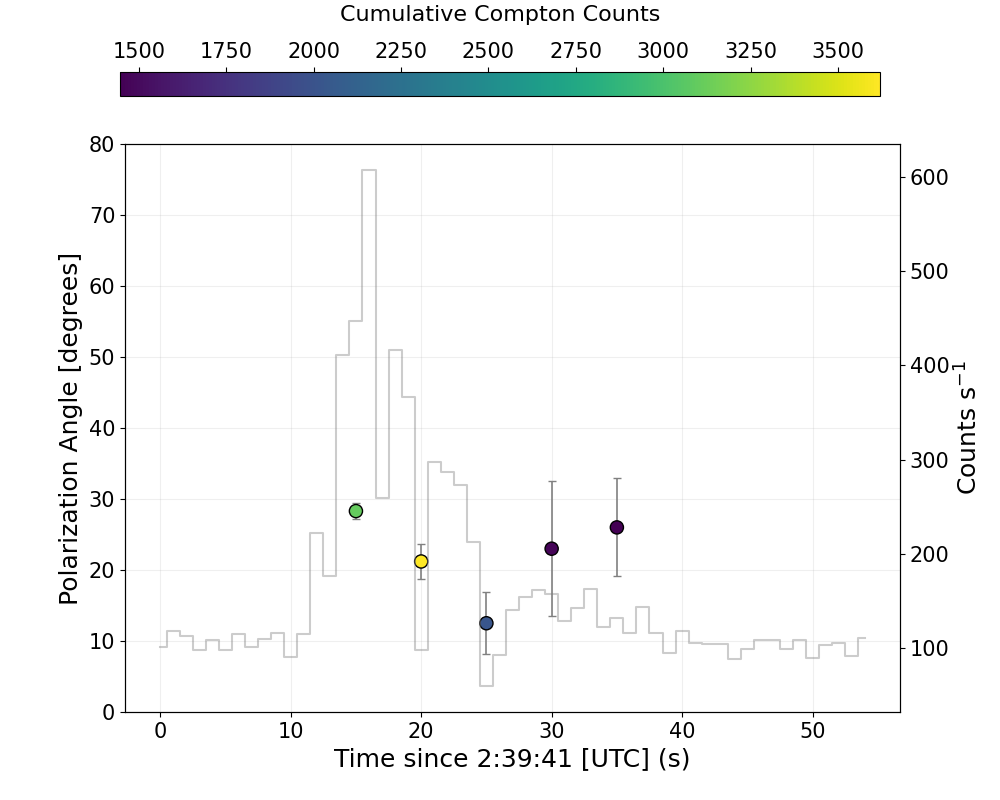}
    \caption{Left panel: The evolution of the polarization degree over time for the second emission episode. Right panel: The evolution of the polarization degree over time for the second emission episode. The color scale represents the cumulative counts. The Compton lightcurve is in the background with light gray steps.}
    \label{fig:GRB_PD_PA_evo}
\end{figure*}

\section{Discussion}
\label{sec:Discussion}
\subsection{Polarization pipeline}

The polarization pipeline used in this paper broadly follows methods outlined in \citet{2009ApJ...695L.208G,2013MNRAS.431.3550G,2014MNRAS.444.2776G}, but it now includes the calibration correction and methodology introduced in \citet{2026A&A...707A.340B}. The latest calibration accounts for INTEGRAL/IBIS instrument degradation and provides a more careful treatment of the Compton counts. We provide further details on the pipeline and on systematic uncertainties in Appendix \ref{app:ibis_polarization_pipeline}.

While the current approach of least-squares already accounts for the correlation of polarization degree and angle \citep{2008ApJ...688L..29F,2010SPIE.7732E..0EW}, a likelihood-based analysis can further account for physical boundaries of polarization parameters and any additional instrumental or background uncertainties \citep{2011APh....34..784K,2017ApJ...848..120L,2026A&A...707A.340B}.  We can further improve the pipeline by incorporating a forward-folding approach similar to the one described in \citet{2019A&A...627A.105B,2020A&A...644A.124K}.
Lastly, the pipeline also does not use a conventional spectral response matrix. Rather, the polarization response accounts for the off-axis angle in the GEANT4 simulation of $a_{100}$, but a dedicated response matrix for $\sim11\degree$ is not utilized in the analysis.

We aim to update the pipeline, present a study based on the entire INTEGRAL/IBIS catalog, and reprocess previously reported GRBs. However, because the data handling for INTEGRAL is quite complex, developing the updated version will take considerable time.

\subsection{Evolution of PD and PA}

The Polarization Degree evolves with time but is roughly consistent within errors. As the number of Compton counts decreases, the error on PD increases as seen from Table \ref{tab:time_res_pol}. In time windows 6 and 7, the Compton counts and S/N ratios decrease considerably; thus, we report only the lower limits. The lower-limit calculation is in line with the method described in \citet{2008ApJ...688L..29F, 2013MNRAS.431.3550G,2014MNRAS.444.2776G}. We also include the contour plots for the lower-limit time windows in the appendix.

Across all time windows, the polarization degree remains high, effectively ruling out photospheric emission in any time bin as theoretical modeling and numerical simulations predict that the polarization degree is lower than 40\% \citep{2009ApJ...698.1042T,2021Galax...9...82G,2014MNRAS.440.3292L,2022ApJ...936...12C,2024ApJ...974..158P,2024ApJ...961..243I}. The evolution of PD is displayed in the left panel of Figure \ref{fig:GRB_PD_PA_evo}. 

The Polarization Angle also shows an evolutionary pattern when we look only at the central values of the polarization angles. The larger error bars for time windows 4 and 7 make it difficult to determine whether the polarization angle truly returns to a higher value, as in time windows 2 and 3, or remains closer to the value in time window 5.

If we take the numbers at face value, the modest evolution of the angle has been attributed to a patchy or turbulent magnetic field. The literature has predicted, through modeling and simulations, that such configurations of magnetic fields lead to the modest evolution of the PA when the jet is non-axisymmetric \citep{2009ApJ...700L.141L,2024MNRAS.52712178G,2024ApJ...961..243I}. Since the evolution is of the order of 20$^\circ$, it is unlikely that this GRB has a globally ordered magnetic field with a helical or toroidal structure \citep{2003ApJ...597..998L,2003ApJ...596L..17G,2020MNRAS.491.3343G}. The results are summarized in the right panel of Figure \ref{fig:GRB_PD_PA_evo}. Moreover, \citet{rybicki1985radiative} derived the dependence of polarization degree on the electron distribution index for a locally ordered magnetic field. They find that for an electron index of 3 in an optically thin region, the maximum degree of polarization is 75\%. While higher electron indices, such as 4 or 5, yield a higher degree of polarization, these indices are not expected in a standard shock scenario, where isotropy is assumed to be similar to \citet{rybicki1985radiative}, and the index range is assumed to be between 2 and 3. 

We can depart from isotropy and assume an anisotropic distribution for electrons, but the standard one-zone synchrotron model no longer applies in this scenario. Rather, the anisotropy arises from magnetic reconnection but can produce the polarization degrees observed for our time-resolved analysis under special conditions such as high magnetization \citep{2023ApJ...959..137C, 2025MNRAS.542..902L}.

\subsection{Evolution of Photon Index}
\begin{figure}[ht]
    \centering
    \includegraphics[width=\columnwidth]{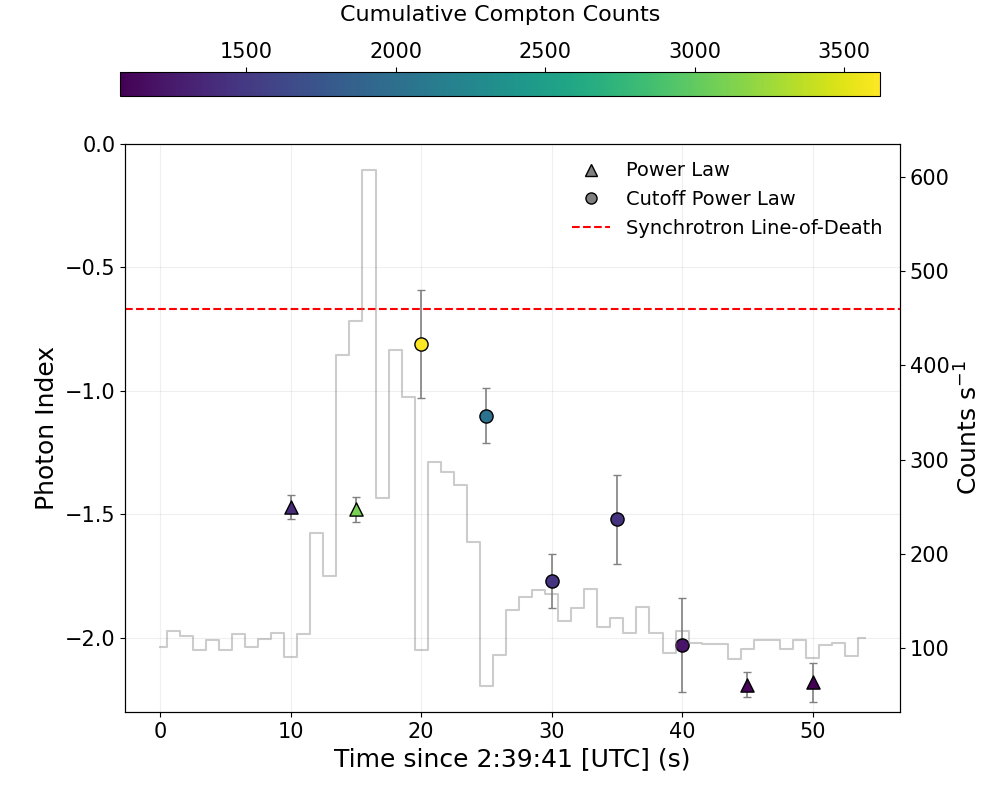}
    \caption{The evolution of the photon index over time for the second emission episode. The triangles represent the bin power law as the best-fit model, while the circles represent the cutoff power law as the best-fit model. Similar to Figure \ref{fig:GRB_PD_PA_evo}, the color scale represents the cumulative counts. The dashed line is the Synchrotron ``line-of-death". }
    \label{fig:Ph_ind_evo}
\end{figure}

We reiterate that the spectral analysis for this GRB is in a different energy band than the polarization analysis. The spectral analysis energy range is 20-200 keV. The data points beyond 200 keV have larger error bars than their values, and thus cannot be used in the analysis. Thus, we cannot accurately capture the GRB's peak energy, as it may lie beyond the 200 keV range. Of the 10 time windows, the GRB was detected in 9 bins. The spectral data across four time bins are best fit by a PL model, whereas the CPL is the best fit for the remaining bins. The evolution of the photon index indicates that it becomes harder near the GRB peak and softer farther from it. We observe a hard-to-soft evolution for the photon index, which does not cross the synchrotron line-of-death (the best-fit photon indices of our models always remain less than -2/3) \citep{1998ApJ...506L..23P} in any time bin as seen in Figure \ref{fig:Ph_ind_evo}. However, since the analysis energy ranges differ for spectral and polarization analysis, we remain conservative in ruling out any models based solely on spectral analysis. Furthermore, we can use the high-energy photon index of the Band-spectrum from the Konus-Wind analysis to estimate the polarization degree. Assuming that the electron distribution remains isotropic and the magnetic field is uniform, the local maximum polarization is given by \citep{rybicki1985radiative},
\begin{equation}
    \Pi_{\max}(\nu)=\frac{\alpha(\nu)+1}{\alpha(\nu)+5/3}
\end{equation}

Note that the spectral index $\alpha=-d\ln F_\nu/d\ln\nu$ in this equation is related to the Band photon index via $-(\beta +1)$. Thus, substituting  $\beta$ as -2.73, we get a polarization degree of $\sim80\%$. However, this is the maximum value for an ordered magnetic field. The net polarization obtained after integrating over the observable region of angular size $1/\Gamma$, which may include more than one incoherent B-field patch, is lower, especially in the case of a multi-patch field. Thus, the measured polarization degree of GRB 181201A remains higher than that expected from theory.

\subsection{Physical Interpretation}
It is important to note that our 10-second time windows have a 5-second overlap. The adjacent data points for the evolution of polarization degree and angle, as well as the spectral index, are thus correlated. Nonetheless, when taken at face value, the results show that the polarization fraction remains roughly consistent across the three time bins, and the central values of the lower limits are not far from those of the first three bins as well. The high polarization degrees effectively strongly disfavor standard photospheric models for this GRB \citep{2014MNRAS.440.3292L,2024ApJ...961..243I}

The polarization angle shows a modest evolution, and a continuous change of $20^\circ$ among the time bins. This evolution leads us to disfavor a single-zone or axisymmetric magnetic field structure \citep{2021Galax...9...82G} as well as a globally ordered toroidal or helical magnetic field. Keeping both the high polarization degree and a modest evolution of the polarization angle in mind, this GRB favors a patchy magnetic field with a non-axisymmetric and possibly a multi-patch emission region \citep{2021Galax...9...82G,2024MNRAS.52712178G}. It is important to note that the high polarization degree that does not change over time is indicative of a ``patch" or a few ``patches" dominating and could have locally coherent magnetic fields.

Folding in the spectral information from IBIS is challenging in this analysis, as the analysis energy ranges differ. Since we may not have cleanly captured the peak energy of this GRB, as the Cadmium Telluride (CdTe) detector is inefficient above 200 keV as the detector becomes transparent, it is hard to tell whether the spectral index remains unchanged after including the high-energy data points. We see that the index does not cross the synchrotron line-of-death, but we cannot rule out any emission mechanism apart from photospheric models. Special cases of Compton drag, such as narrow jets which are viewed close to their sharp edge,  synchrotron mechanism with multi-zone emission \citep{2004MNRAS.347L...1L, 2009ApJ...698.1042T}, and the Internal-Collision-Induced Magnetic Reconnection and Turbulence (ICMART) still remain viable \citep{2011ApJ...726...90Z}.

\section{Conclusion}
\label{sec:Conclusion}
Analyzing all GRBs within
7$\degree$ of the instrument boresight and including any GRB that lie beyond 7$\degree$ with high S/N in the Compton count space, we find one additional polarized GRB in the INTEGRAL/IBIS catalog. This is in addition to the three GRBs previously reported by the INTEGRAL team \citep{2009ApJ...695L.208G,2011MNRAS.413.2173G,2014MNRAS.444.2776G}. 

GRB 181201A shows a high degree of polarization in both the time-integrated and time-resolved cases. The PD remains high over multiple time bins, and PA shows a weak temporal evolution. The photon index hardens near the burst peak and softens in other time windows. The index in any window does not cross the synchrotron line-of-death. However, the difference in the energy range of the polarization and spectral analysis prevents us from cleanly combining the information.

Using the polarization information as the primary indicator, we strongly disfavor the photospheric emission mechanism model for this GRB. The modest evolution of the PA, is indicative of a patchy magnetic field with a non-axisymmetric jet and a multi-patch emission region \citep{2021Galax...9...82G,2024MNRAS.52712178G}. Due to a lack of spectral information beyond 200 keV, and solely based on the polarization values, the Compton drag mechanism and a synchrotron emission mechanism with a multi-zone emission \citep{2004MNRAS.347L...1L, 2009ApJ...698.1042T} remain viable for GRB 181201A.
We want to highlight that special conditions for Compton Drag, where the jet is narrow and is viewed close to the sharp edge, can produce such a high polarization degree. The standard one-zone synchrotron emission with an isotropic electron distribution is disfavored for this GRB.

This analysis reiterates the importance of time-resolved spectropolarimetry as the results allow us to rule out certain emission mechanisms and magnetic field structures. The upcoming instruments, such as POLAR-2 \citep{2021SPIE11444E..2VH}, COSI \citep{2019BAAS...51g..98T}, and Daksha \citep{2024ExA....57...24B}, are vital to GRB science, as spectropolarimetry offers a unique window into the properties of GRB progenitors. Missions beyond those listed here would benefit from designing instruments that enable time-resolved analysis for GRBs, as they remain a crucial probe into GRB physics.

Finally, an updated analysis with an improved spectropolarimetric pipeline is in the works. The pipeline will incorporate Compton mode data for the spectral analysis beyond 200 keV as demonstrated in \citet{2026A&A...707A.340B} for Crab. We aim to analyze all GRBs in the INTEGRAL/IBIS archive with an improved understanding of systematic errors. While the absolute values of the polarization degree could change, we expect the temporal trend to be retained. Thus, the values reported by the current version of the pipeline, which capture the temporal evolution, still remain important.

\section{Acknowledgements}

We thank the anonymous referee for their suggestions that have helped in significantly improving the paper. We acknowledge Clemson University for its support with the license for the IDL software. D.S.B. and N.S.K. acknowledge that this research was supported in part by the Clemson University Doctoral Dissertation Completion Grant.

\software{This work benefited from the following software: \textsc{IDL}\citep{idl_software}, \textsc{Matplotlib} \citep{Hunter:2007}, and \textsc{Plotly} \citep{plotly}.}

\bibliography{INTEGRAL_GRB_Polarization_analysis_arXiv}{}
\bibliographystyle{aasjournalv7}

\section*{Appendix}
\subsection{Polarization pipeline}

\label{app:ibis_polarization_pipeline}

The design of INTEGRAL/IBIS which has two detector planes, ISGRI and PICsIT, allows us to use it as a polarimeter for prompt GRB polarization measurements. Since the prompt emission is dominated by X-rays and gamma-rays, Compton scattering in the detector is the underlying physical phenomenon that allows us to build the polarigrams. The method was originally developed by \citet{2008ApJ...688L..29F} and was eventually applied to GRBs by \citet{2009ApJ...695L.208G,2013MNRAS.431.3550G, 2014MNRAS.444.2776G}. This pipeline further includes the recent in-orbit calibration prescribed by \citet{2026A&A...707A.340B}.

The energy and direction of the photon are reconstructed as it first scatter from the top layer (ISGRI) and is then absorbed by the lower layer (PICsIT). This interaction is then used to determine the energy and the azimuthal angle of the photon as 

\begin{equation}
\frac{{\rm d}\sigma}{{\rm d}\Omega}
=
\frac{r_{0}^{2}}{2}
\left(\frac{E'}{E}\right)^{2}
\left[
\frac{E'}{E}
+
\frac{E}{E'}
-
sin^2\theta-\sin^{2}\theta\cos2\phi
\right],
\label{eq:polarized_KN}
\end{equation}
where 

\begin{equation}
E'
=
\frac{E}
{1+\dfrac{E}{m_{\rm e}c^{2}}(1-\cos\theta)}.
\label{eq:compton_energy}
\end{equation}

This is because, in linear polarization, photons scatter preferentially perpendicular to the incident electric field. Furthermore, handling of the chance coincidences, event-selection criterion and energy calibration follow from \citet{2026A&A...707A.340B}. The photons are divided into six azimuthal bins of 30$^\circ$ each. For each bin, the GRB flux is obtained by deconvolving the detector shadowgrams through the coded aperture mask. The corrected Compton flux is written as  

\begin{equation} C_{\rm t}(E,t) = \alpha(E,t)\,C_{\rm r}(E,t) - \beta(t)\,S(E,t), \label{eq:compton_correction} \end{equation} 

The resulting polarigram is then fitted with

\begin{equation} N(\phi)=S[1+a_0\ \cos\ 2(\phi-\phi_0)], \qquad \rm PD=\frac{a_{0}}{a_{100}}, \qquad {\rm PA}=\phi_0-\pi/2+n\pi, \label{eq:ibis_polarigram} \end{equation} 
where $N$ is the number of detected Compton events as a function of the azimuthal angle $\phi$, $S$ is the normalization, $a_0$ is the modulation amplitude, and $\phi_0$ is the direction of the maximum scattering. Thus, we are not yet using a POLAR-like forward-folding methodology in the INTEGRAL pipeline.

For GRB 181201A, which was detected at an off-axis angle of $\sim11^{\circ}$, we use a dedicated GEANT4 simulation for a$_{100}$ using the burst spectrum and the off-axis angle. The a$_{100}$ was subject to the same selection criteria as the observations, ensuring that the polarization response is energy-dependent. Thus, the estimated systematic uncertainty on its value is about 5\%. Finally, it is worth noting that the adjacent 10-second intervals have a 5-second overlap, and thus they are not statistically independent in polarization or spectral analyses.

\subsection{Polarization degree lower-limits}
\label{app:lower-limits}

\begin{figure*}[ht!]
    \centering
    \includegraphics[width=0.45\textwidth]{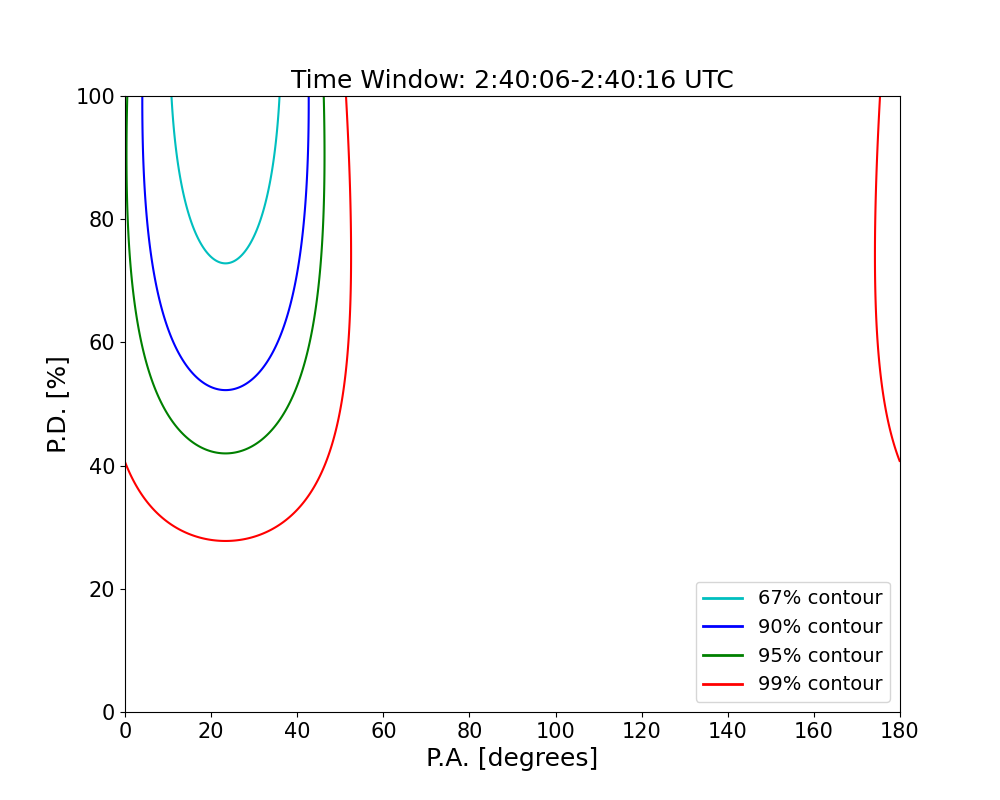}
    \includegraphics[width=0.45\textwidth]{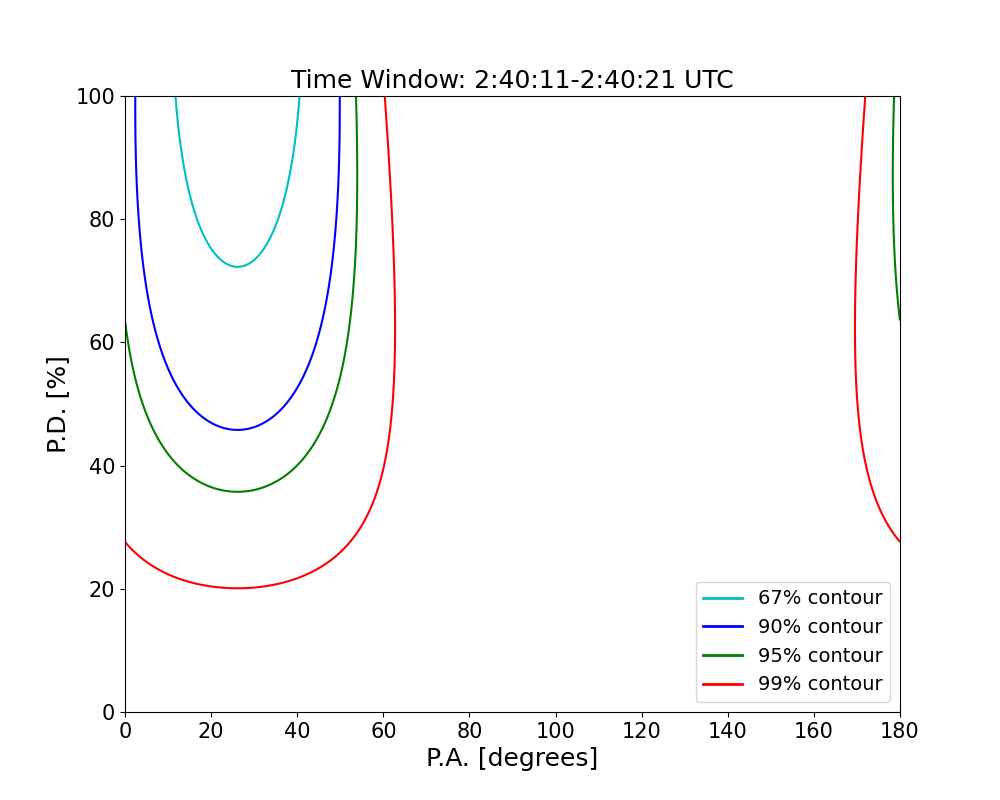}
    \caption{Left panel: The contours for the lower limit for time window 6 of GRB 181201A. Right panel: The contours for the lower limit for time window 7 of GRB 181201A.}
    \label{fig:lower-limits}
\end{figure*}

To calculate the lower-limit values for time windows 6 and 7, we utilize the procedure outlined in \citet{2008ApJ...688L..29F,2013MNRAS.431.3550G,2014MNRAS.444.2776G}. For GRBs 061122 and 140206A, the authors jointly fit the polarization degree and angle space to determine the lower limits. We report the 90\% confidence interval values. The contours for different confidence intervals are illustrated in Figure \ref{fig:lower-limits}. 

\subsection{Time-integrated and resolved analysis}
\begin{deluxetable}{cccccc}[!ht]
\tablecaption{Time-integrated and individual peak polarization analysis} \label{tab:time_int_pol}
\tablehead{
\colhead{T$_{\rm start}$ } & \colhead{T$_{\rm stop}$ } & \colhead{Counts} & \colhead{S/N} & \colhead{PD} & \colhead{PA} }
\startdata
2:38:00 & 2:40:36 & 20261 & 28.9 & 91 $\pm$ 5 & 22 $\pm$ 1.5\\
2:38:00 & 2:38:40 & 4535 & 2.4 & - & - \\
2:39:41 & 2:40:36 & 9331 & 35.9 & 93 $\pm$ 8 & 22.5 $\pm$ 2.4\\
\hline
\enddata
\tablecomments{The time bins for the polarimetric and the spectral analysis for GRB 181201A. The times in the first two columns are in UTC. The Counts are the number of Compton events detected in that interval. The last two columns are the Polarization degree in \% and the Polarization angle in degrees, respectively.}

\end{deluxetable}

\begin{deluxetable*}{ccccccccc}[!hb]
\tablecaption{Time-integrated spectral analysis} \label{tab:time_int_spec}
\tablehead{
\colhead{\shortstack{T$_{start}$ \\[0.5ex] (UTC)} } & \colhead{\shortstack{T$_{stop}$ \\[0.1ex]  (UTC)}} & \colhead{\shortstack{Model \\[2.3ex] ~}} & \colhead{\shortstack{A \\[0.6ex] ($\mathrm{ph/cm^2/s/keV}$)}} & \colhead{\shortstack{$\alpha$ \\[2.6ex] ~}} & \colhead{\shortstack{E$_{0}$ \\[0.5ex] (keV)}} & \colhead{\shortstack{E$_{peak}$ \\ [0.3ex]  (keV)}}& \colhead{\shortstack{$\chi^2$/dof \\[2.0ex] ~}} & \colhead{\shortstack{AIC$_c$ \\[2.1ex] ~}}}

\startdata
2:38:00 & 2:40:36 & CPL & 12.6$\pm$2.9 & 1.6 $\pm $0.07 & 263.9 $\pm$ 87.2 & 105.56 & 50.1/45 & 57.16\\
2:38:00 & 2:38:40 & CPL & 2.8$\pm$1.2 & 1.2 $\pm $0.1 & 159.2 $\pm$ 53.2 & 127.36 & 53.44/45 & 59.98\\
2:39:41 & 2:40:36 & CPL & 53.7$\pm$12.3 & 1.75 $\pm $0.08 & 316.2 $\pm$ 102.1 & 94.86 & 62/45 & 68.54\\
\hline
\enddata
\tablecomments{The times in the first two columns are in UTC. The second column indicates the best-fit model. The remaining columns are the spectral indices and the characteristic energies for the best-fit models.}

\end{deluxetable*}

\begin{deluxetable*}{cccccc}
\tablecaption{Time-resolved polarization analysis} \label{tab:time_res_pol}
\tablehead{
\colhead{T$_{start}$ } & \colhead{T$_{stop}$ } & \colhead{Counts} & \colhead{S/N} & \colhead{PD} & \colhead{PA} }
\startdata
2:39:41 & 2:39:51 & 1121 & 1 & - & - \\
2:39:46 & 2:39:56 & 1389 & 7.9 & - & -\\
2:39:51 & 2:40:01 & 3102 & 30.6 & 83 $\pm$ 3 & 28.3 $\pm$ 1.1\\
2:39:56 & 2:40:06 & 3620 & 35.9 & 81 $\pm$ 7 & 21.2 $\pm$ 2.5\\
2:40:01 & 2:40:11 & 2025 & 21 & 87 $\pm$ 13 & 12.5 $\pm$ 4.4\\
2:40:06 & 2:40:16 & 1470 & 11.9 & $>75\%$ & 23 $\pm$ 9.5\\
2:40:11 & 2:40:21 & 1446 & 9.6 & $>74\%$ & 26 $\pm$ 6.9\\
2:40:16 & 2:40:26 & 1215 & 5.2 & - & -\\
2:40:21 & 2:40:31 & 1111 & 1.9 & - & -\\
2:40:26 & 2:40:36 & 1078 & 2.1 & - & -\\
\hline
\enddata
\tablecomments{The time bins for the polarimetric and the spectral analysis for GRB 181201A. The times in the first two columns are in UTC. The Counts are the number of Compton events detected in that interval. The last two columns are the Polarization degree in \% and the Polarization angle in degrees, respectively.}

\end{deluxetable*}

\begin{deluxetable*}{ccccccccc}
\tablecaption{Time-resolved spectral analysis} \label{tab:time_res_spec}
\tablehead{
\colhead{\shortstack{T$_{start}$ \\[0.5ex] (UTC)} } & \colhead{\shortstack{T$_{stop}$ \\[0.1ex]  (UTC)}} & \colhead{\shortstack{Model \\[2.3ex] ~}} & \colhead{\shortstack{A \\[0.6ex] ($\mathrm{ph/cm^2/s/keV}$)}} & \colhead{\shortstack{$\alpha$ \\[2.6ex] ~}} & \colhead{\shortstack{E$_{0}$ \\[0.5ex] (keV)}} & \colhead{\shortstack{E$_{peak}$ \\ [0.3ex]  (keV)}}& \colhead{\shortstack{$\chi^2$/dof \\[2.0ex] ~}} & \colhead{\shortstack{AIC$_c$ \\[2.1ex] ~}}}

\startdata
2:39:41 & 2:39:51 & - & - & - & - & - & - & -\\
2:39:46 & 2:39:56 & PL & 11.4$\pm$2.4 & 1.47 $\pm $0.05 & - & - & 59.1/46 & 63.38\\
2:39:51 & 2:40:01 & PL & 25.1$\pm$4.8 & 1.48 $\pm $0.05 & - & - & 49.6/46 & 53.88\\
2:39:56 & 2:40:06 & CPL & 27.2$\pm$19.7 & 0.81 $\pm $0.22 & 75.4 $\pm$ 19.9 & 89.73 & 47.51/45 & 50.1\\
2:40:01 & 2:40:11 & CPL & 206.4$\pm$68.4 & 1.61 $\pm $0.11 & 239.3 $\pm$ 91.1 & 93.3 & 57.7/45 & 64.3\\
2:40:06 & 2:40:16 & CPL & 225.3$\pm$74.5 & 1.77 $\pm $0.11 & 236.2 $\pm$ 94.1 & 54.33 & 49.9/45 & 56.4\\
2:40:11 & 2:40:21 & CPL & 65.6$\pm$32.6 & 1.52 $\pm $0.18 & 98.6 $\pm$ 27.2 & 47.3 & 49.2/45 & 55.73\\
2:40:16 & 2:40:26 & CPL & 158.5$\pm$83.3 & 2.03 $\pm $0.19 & 306.5 $\pm$ 235.2 & - & 63.9/45 & 70.5\\
2:40:21 & 2:40:31 & PL & 183.4$\pm$37.4 & 2.19 $\pm $0.05 & - & - & 45.7/46 & 50.1\\
2:39:26 & 2:40:36 & PL & 91.8$\pm$29.1 & 2.18 $\pm $0.08 & - & - & 71/46 & 75.2\\
\hline
\enddata
\tablecomments{The times in the first two columns are in UTC. The second column indicates the best-fit model. The remaining columns are the spectral indices and the characteristic energies for the best-fit models.}

\end{deluxetable*}

\end{document}